# Artificial-Intelligence-Based Hybrid Extended Phase Shift Modulation for the Dual Active Bridge Converter with Full ZVS Range and Optimal Efficiency

Xinze Li, *Student Member, IEEE*, Xin Zhang, *Senior Member, IEEE*, Fanfan Lin, *Student Member, IEEE*, Changjiang Sun, *Member, IEEE*, Kezhi Mao, *Member, IEEE*.

*Abstract*—Dual active bridge (DAB) converter is the key enabler in many popular applications such as wireless charging, electric vehicle and renewable energy. ZVS range and efficiency are two significant performance indicators for DAB converter. To obtain the desired ZVS and efficiency performance, modulation should be carefully designed. Hybrid modulation considers several single modulation strategies to achieve good comprehensive performance. Conventionally, to design a hybrid modulation, harmonic approach or piecewise approach is used, but they suffer from time-consuming model building process and inaccuracy. Therefore, an artificial-intelligence-based hybrid extended phase shift (HEPS) modulation is proposed. Generally, the HEPS modulation is developed in an automated fashion, which alleviates cumbersome model building process while keeping high model accuracy. In HEPS modulation, two EPS strategies are considered to realize optimal efficiency with full ZVS operation over entire operating ranges. Specifically, to build data-driven models of ZVS and efficiency performance, extreme gradient boosting (XGBoost), which is a state-of-the-art ensemble learning algorithm, is adopted. Afterwards, particle swarm optimization with state-based adaptive velocity limit (PSO-SAVL) is utilized to select the best EPS strategy and optimize modulation parameters. With 1 kW hardware experiments, the feasibility of HEPS has been verified, achieving optimal efficiency with maximum of 97.1% and full-range ZVS operation.

## I. INTRODUCTION

The dual active bridge (DAB) isolated dc-dc converter, since its first appearance in 1990s, has incrementally gained research attentions and popularity in industry as being benefited from its high power density, the capability of bidirectional power transfer and its wide zero voltage switching (ZVS) range [1]. DAB converter consists of two full bridges and one high-frequency transformer in between, which enjoys more compact size and lighter weight than bulky line-frequency transformer. This topology has omnipresent applications nowadays such as, electric vehicle [2], wireless charging [3], solar panel [4], solid state transformer [5], etc.

To actively control power flow and improve operating performance, modulations of DAB converters have been widely studied, among which the phase shift modulation is commonly adopted for its easy implementation. Single phase shift (SPS) modulation has one degree of control freedom, which is the outer phase shift between two bridges [5]. Although SPS modulation is a simple strategy, it suffers from narrow ZVS range and low efficiency in light load conditions. To achieve wider ZVS range and better efficiency, extended phase shift (EPS) modulation and dual phase shift (DPS) modulation consider one more degree of control freedom, which is the inner phase shift of full bridges. DPS modulation applies the same inner phase shift to both bridges [7]. EPS modulation applies the inner phase shift to only one bridge, so there are two EPS strategies depending on which bridge applies the inner phase shift [8]. However, EPS or DPS alone still cannot achieve full ZVS range and optimal efficiency [9]. With the two inner phase shifts and the outer phase shift to be independently tunable, triple phase shift (TPS) modulation has three degrees of control freedom [10]. TPS is the generalized and improved version of SPS, EPS and DPS, but it is complicated to implement. Besides single modulation strategies, some research works focus on hybrid modulation (HM) to obtain good holistic performance over wide operating ranges. For instance, Amit *et al.* analyzed different modulation modes of TPS to realize optimal efficiency over wide operating conditions [11]. Shen *et al.* combined EPS and several modulation modes of TPS for full soft switching range with minimized conduction loss [12]. Deng *et al.* adjusted phase shift and pulse duty cycle simultaneously to achieve good efficiency over wide voltage range and extend ZVS range [13].

With regards to the optimization objective of modulations for DAB converters, ZVS range is a desirable target to be optimized. Wide ZVS range is of great significance. First, realization of ZVS can reduce the switching losses, which are dominant when the switching frequency is high. Second, full ZVS operation can alleviate many undesired phenomena like voltage polarity reversal, voltage sag and phase drift [14]. Moreover, soft switching operation can reduce electromagnetic interference, protecting the control circuit [15]. Consequently, in this article, an HM consisting of two EPS strategies is optimized to realize full ZVS range under all operating conditions. Besides, power transfer efficiency is also considered as an optimization target for a good comprehensive performance.

Nonetheless, the conventional modeling approaches of ZVS and efficiency are truly cumbersome, and the deduced model may suffer from inaccuracy. For example, traditionally, ZVS model is built with the piecewise method [9], [16], [17] or the harmonic method [18], [19]. The piecewise method requires segment-by-segment analysis of current and voltage for all operating modes in each single modulation. The harmonic model may fail to reach high accuracy if limited harmonic components are considered for the sake of simplicity. In HM, the time-consuming and inaccurate problems will even aggravate as more modulation strategies have to be analyzed. Therefore, how to conquer the challenges of burdensome model building process of ZVS and efficiency and low model accuracy is a major concern of this article.

As artificial intelligence (AI) thrives, more researchers in power electronics start to leverage the advanced AI tools to relieve the burden of engineers and improve design accuracy. For instance, Dragičević *et al.* modeled the reliability of converters using two neural networks (NN) to design converters with predetermined lifetime [20]. Li *et al.* adopted batch-normalization NN to compute power losses for optimizing synchronous buck converter in EV [21]. NN was also utilized to model current stress [10] and efficiency [22] of DAB converter under TPS modulation. Reinforcement learning was applied to optimize reactive power [23].

Nevertheless, these AI techniques can be problematic. If the learnable parameters of AI models are insufficient, such as support vector machine, decision tree and shallow NN, low accuracy problem may occur [21]. Besides, if deep learning model is used to provide stronger learning capability, it may potentially suffer from overfitting and heavy computation problems. An AI technique that achieves a balance between high accuracy and fast computational speed is required. Luckily, an AI technique in ensemble learning, extreme gradient boosting (XGBoost), serves as a good solution. XGBoost utilizes boosting framework to learn a bag of weak models in sequence: the next model adapts to the error of previous models [24]. With the weak models ensembled altogether, XGBoost can give highly accurate predictions, and its learning capability can increase by stacking more base models. Meanwhile, cache optimization and parallelization features of XGBoost make it a fast-speed algorithm [24]. In addition, the structure of XGBoost model is easy to optimize because only few hyperparameters are adjustable. In many international AI competitions like Kaggle, XGBoost earns its reputation for its superb performance.

In this article, utilizing the state-of-the-art AI techniques, a hybrid EPS (HEPS) modulation with full ZVS range and optimal efficiency for the entire voltage and power ranges is proposed. This hybrid modulation is composed of two EPS strategies from the simultaneous considerations of good operating performance and ease of implementation. Generally, HEPS consists of two stages. In Stage I, data-driven models of ZVS and efficiency are built. With the data generated by simulations, XGBoost algorithm is adopted to automatically build surrogate models for ZVS and efficiency. In Stage II, the best modulation strategy is selected, and modulation parameters are optimized to achieve full ZVS range with optimal efficiency. Stage II utilizes the cutting-edge particle swarm optimization with state-based adaptive velocity limit (PSO-SAVL) to optimize efficiency with the constraint of all-switch full ZVS range. The proposed HEPS contributes two main points to the existing research works: first, optimal efficiency under full ZVS range is realized in the proposed HEPS modulation; second, it significantly reduces the amount of manpower involved and achieves high modeling accuracy.

The structure of this paper is discussed as follows. In Section II, the operating principles of the two considered EPS strategies are introduced. Section III illustrates the potential problems in conventional modeling approaches of ZVS and efficiency. Stage I and Stage II of the proposed HEPS are illustrated in detail in Section IV and V, respectively. An easy-to-follow step-by-step case study is given in Section VI. In Section VII, hardware experimental results are shown and analyzed. Section VIII summarizes the conclusion of this article.

## II. PRELIMINARY: BASICS ABOUT EPS MODULATIONS

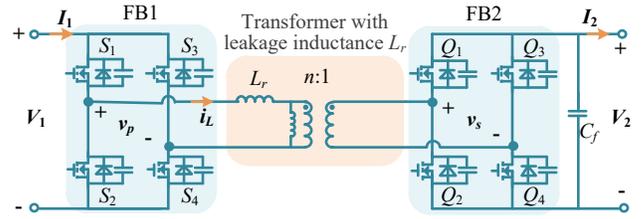

Fig. 1. Topology of a non-resonant DAB converter with leakage inductance $L_r$.

The non-resonant DAB converter with single leakage inductance $L_r$ in between is shown in Fig. 1. It consists of two bridges FB1 and FB2, and a galvanically isolated transformer. The turn ratio of transformer is $n$:1. $V_1$, $I_1$ and $V_2$, $I_2$ are the dc-side voltage and current of primary bridge and secondary bridge, respectively. $v_p$ and $v_s$ are the ac-side voltages of primary and secondary bridges. $i_L$ is the power transfer current through $L_r$.

### A. Fundamentals of EPS1

Generally, there are two adjustable control parameters in EPS modulations: outer phase shift $D_o$ which controls power transfer, and inner phase shift $D_{in}$ which adjusts the shape of inductor current $i_L$ to improve operating performance. With adjustable $D_{in}$, a zero-level voltage plateau is introduced in either primary bridge FB1 or secondary bridge FB2. Attributable to the zero-level voltage plateaus, backflow power and circulating current can be largely reduced if $D_{in}$ is properly adjusted [13].

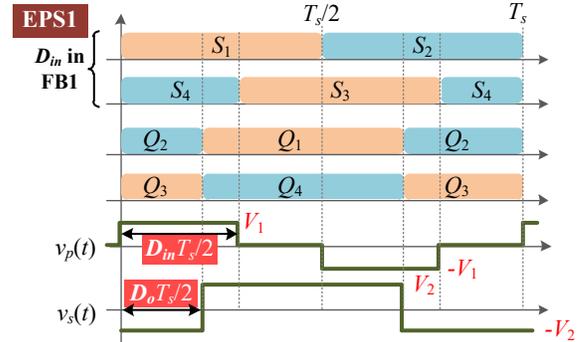

Fig. 2. Fundamentals of EPS1.

As the fundamentals shown in Fig. 2, since the inner phase shift $D_{in}$ is applied to FB1, this EPS strategy is named as EPS1. With EPS1, the driving signal for $S_3$ has $D_{in}T_s/2$ time delay than that for $S_1$, while the driving signals for $Q_1$ and $Q_3$ are complementary with 50% duty ratio. The value of $D_{in}$ lies within [0, 1]. When $D_{in} = 1$, no zero-level voltage is introduced, and thus $v_p$ will still be a two-level wave. The outer phase shift $D_o$ regulates the time delay ($D_oT_s/2$) between $Q_1$ and $S_1$. $D_o$ has the range of [-1, 1], where the maximum forward power is achieved at $D_o = 0.5$, and no power is transferred when $D_o = 0$ [16]. In EPS1, $v_p$ is a three-level ac voltage wave, while $v_s$ is still a two-level ac square wave.

## B. Fundamentals of EPS2

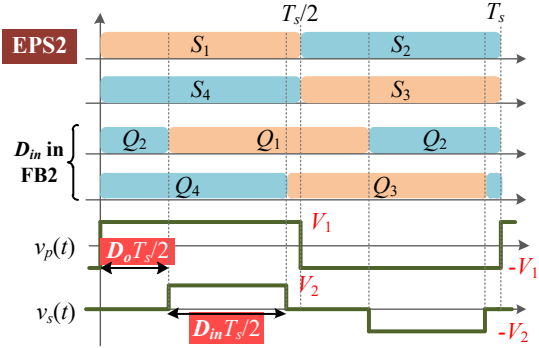

Fig. 3. Fundamentals of EPS2.

The difference between EPS2 and EPS1 is that EPS2 adjusts the inner phase shift $D_{in}$ of FB2 rather than that of FB1, as shown in Fig. 3. With EPS2, the driving signals for $S_1$ and $S_3$ are complementary square waves with 50% duty ratio. Whereas, because $D_{in}$ is applied to FB2, there exists a time delay between the driving signals of $Q_1$ and $Q_3$. The definition of $D_o$ is the same as EPS1. The adjustable ranges of $D_o$ and $D_{in}$ in EPS2 are [-1, 1] and [0, 1], respectively. Conversely, with EPS2, $v_p$ is a two-level waveform, while $v_s$ is a three-level waveform.

In comparison to SPS, benefiting from the extra zero-level voltage plateau, EPS strategy has higher efficiency, lower current stress, and broader ZVS range [17]. With the two EPS strategies considered at the same time, $D_{in}$ can be applied to both full bridges, making it possible for all-switch ZVS over full operating ranges. In this article, through properly shifting between EPS1 and EPS2 and adjusting $D_{in}$, optimal efficiency with full ZVS operation can be realized.

## III. CONVENTIONAL MODELING APPROACHES OF ZVS AND EFFICIENCY

This section discusses the two main modeling approaches of ZVS and efficiency for DAB converters under phase shift modulations in the existing literatures: the piecewise approach and the harmonic approach. Section III-A and Section III-B illustrate the piecewise approach and the harmonic approach for modeling ZVS, respectively. Section III-C discusses the piecewise and harmonic approaches for modeling efficiency.

### A. Piecewise Approach for Modeling ZVS

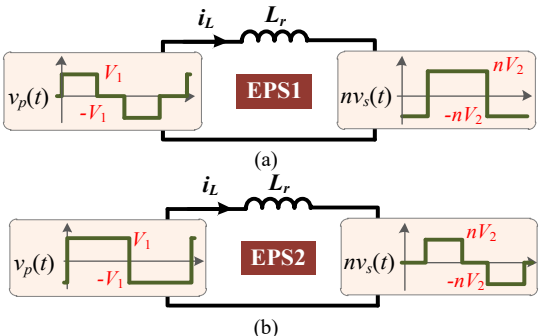

Fig. 4. Equivalent circuits of DAB converter under: (a) EPS1; (b) EPS2.

The equivalent circuits of DAB converter operating under EPS1 and EPS2 are shown in Fig. 4 (a) and (b), respectively. Both piecewise and harmonic approaches are based on the equivalent circuits in Fig. 4.

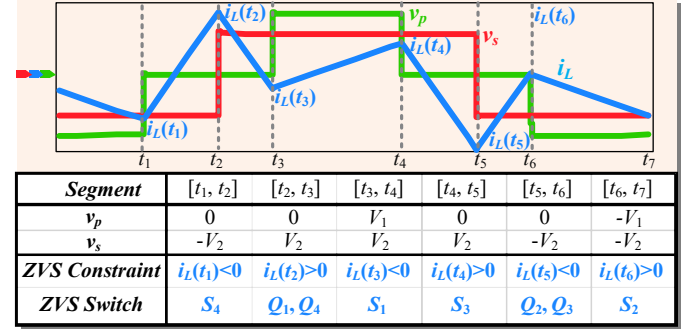

Fig. 5. Example of piecewise approach for modeling ZVS.

The piecewise approach for analyzing ZVS under EPS1 is given as an example in Fig. 5. Piecewise approach analyzes the inductor current $i_L$ segment-by-segment through applying the volt-second balance principle [25]. The time-variant expressions $i_L(t)$ of all operating modes of EPS1 and EPS2 should be deduced. After that, to achieve all-switch ZVS operation, the values of $i_L$ at the time of commutation have to follow the six constraints in Fig. 5. These constraints ensure that the drain-source voltage of switch is clamped to zero by antiparallel diode when the switch commutates [18]. Furthermore, these steps have to be repeated multiple times to achieve all-switch ZVS over entire operating ranges.

Considering the high non-linearity of $i_L$ and the requirements to meet all ZVS constraints over full operating ranges, it would be extremely complex, error-prone, and time-consuming to build the ZVS model with piecewise approach.

### B. Harmonic Approach for Modeling ZVS

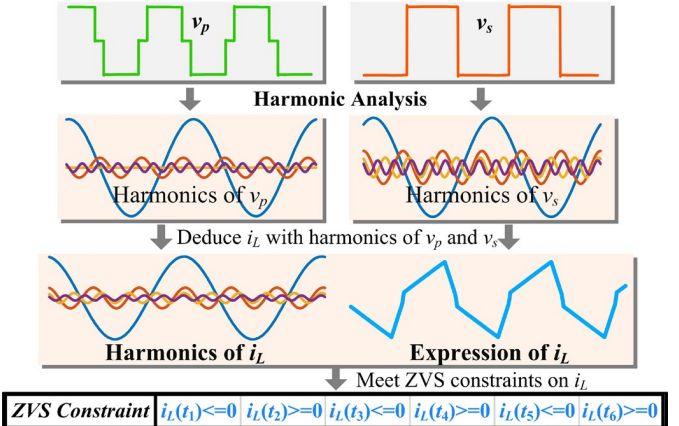

Fig. 6. Example of harmonic approach for modeling ZVS.

Fig. 6 gives an example of applying harmonic approach to model ZVS under EPS1. First, ac-side voltage waves $v_p$ and $v_s$ will be converted into harmonics in Fourier series as shown in (1) and (2) [18], [26], in which $\varphi_o$ is the outer phase shift angle, and $\varphi_{i1}$ and $\varphi_{i2}$ are the inner phase shift angles for FB1 and FB2, respectively. Harmonic components of $i_L$ are then deduced by dividing $v_p$-$v_s$ by the impedance of $L_r$, and thus the time-variant formula of $i_L$ is also expressed in the form of Fourier series. To realize all-switch ZVS operation, the constraints on $i_L$ as listed in Fig. 6 should be satisfied.

$$v_p(t) = \sum_{k=1,3,\ldots,\infty} \frac{4V_1}{k\pi} \cos\left(k\frac{\varphi_{i1}}{2}\right) \sin(k\omega_0 t) \quad (1)$$

$$v_s(t) = \sum_{k=1,3,\ldots,\infty} \frac{4V_2}{k\pi} \cos\left(k\frac{\varphi_{i2}}{2}\right) \sin(k(\omega_0 t - \varphi_o)) \quad (2)$$

The harmonic approach for modeling ZVS suffers from the compromise between high model accuracy and fast computation. If all harmonic components are considered, it suffers from slow computation. And if only fundamental component is used, accuracy is largely sacrificed. If limited components are taken into account, the modeling of $i_L$ through harmonic analysis may be easier than piecewise approach, but the later steps to satisfy all ZVS constraints over full operating ranges remain tedious.

### C. Piecewise and Harmonic Approaches for Modeling Efficiency

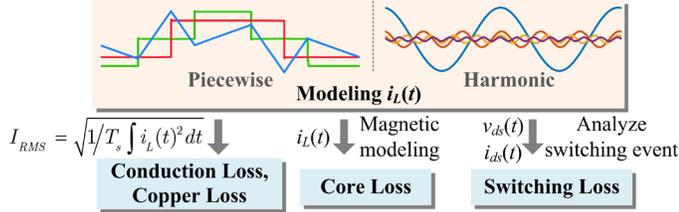

Fig. 7. Modeling efficiency.

To build models of efficiency for DAB converter under EPS modulations, the expressions of inductor current $i_L(t)$ should be obtained first with either piecewise approach in Section III-A or harmonic approach in Section III-B. Based on $i_L(t)$, $I_{RMS}$ is then computed with the square root of the integration of $i_L(t)^2$ to analyze conduction and copper losses. The core loss is analyzed through magnetic modeling with $i_L(t)$. Switching loss is obtained by analyzing drain-source voltage $v_{ds}(t)$ and drain-source current $i_{ds}(t)$ during commutation.

In a nutshell, to build models of ZVS and efficiency for DAB converter under EPS modulations, there are two main problems: cumbersome and time-consuming analysis process, and inaccuracy due to mathematical approximations. Moreover, high non-linearity of ZVS and efficiency models makes it challenging to optimize efficiency with the constraints of all-switch ZVS over entire operating ranges by hand. To tackle these challenges, an AI-based HEPS modulation is proposed in this article.

## IV. STAGE I OF THE PROPOSED HEPS MODULATION: BUILD DATA-DRIVEN MODELS OF ZVS AND EFFICIENCY

In this section, Stage I of the proposed HEPS modulation which builds the data-driven models of ZVS and efficiency for DAB converter under EPS1 and EPS2 is discussed.

### A. Stage I: Build Data-Driven Models of ZVS and Efficiency with XGBoost Algorithm

As a preparation of Stage I, the operating specifications should be predetermined, including switching frequency $f_s$, dead time, range of output power $[P_{min}, P_{max}]$, input voltage $V_1$ and range of output voltage $[V_{2,min}, V_{2,max}]$. The value of leakage inductance $L_r$ is designed with (3) considering maximum power transfer [10].

$$L_r \leq \frac{nV_1V_{2\min}}{8f_sP_{\max}} \quad (3)$$

The detailed flowchart of Stage I is shown in Fig. 8, consisting of three main steps. First, combinations of operating and modulation parameters ($P$, $V_2$, $S$, $D_{in}$) should be selected for running simulations. $S$ is the modulation selector: $S = 0$ and $S = 1$ represent EPS1 and EPS2, respectively. The numbers of samples of $P$, $V_2$ and $D_{in}$ are $N_1$, $N_2$ and $M_1$, which are uniformly selected from the ranges of $[P_{min}, P_{max}]$, $[V_{2,min}, V_{2,max}]$ and $[0, 1]$, respectively. Consequently, the total number of combinations for simulation will be $2 \times N_1 \times N_2 \times M_1$. The uniform sampling ensures a nicely covered data distribution for the benefit of XGBoost learning.

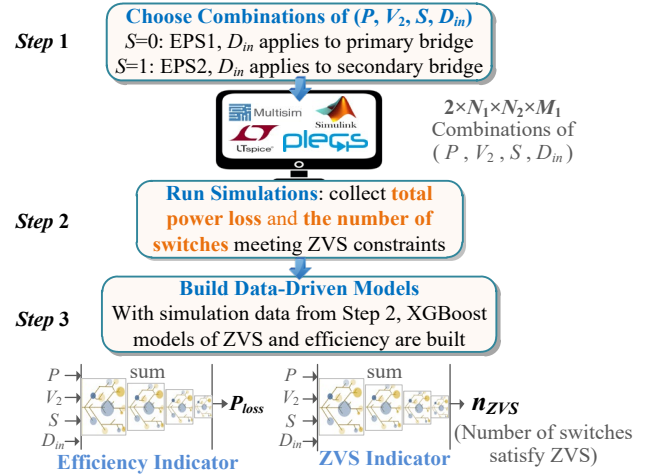

Fig. 8. Flowchart of Stage I of the proposed AI-based HEPS modulation.

Step 2 conducts the simulations for the selected $2 \times N_1 \times N_2 \times M_1$ combinations of operating and modulation parameters. In this article, PLECS simulation is used for its fast speed and high accuracy. In the simulations, total power loss $P_{loss}$ and the number of switches that satisfy ZVS constraints $n_{ZVS}$ are collected as the training targets for Step 3.

In Step 3, trained on the data from Step 2, two data-driven models of ZVS and efficiency are automatically built with XGBoost algorithm. The inputs of both XGBoost models are $P$, $V_2$, $S$ and $D_{in}$. The outputs of the two models are total power loss $P_{loss}$ and the number of switches satisfying ZVS constraints $n_{ZVS}$, respectively. The collected simulation data is partitioned into training set (70%), validation set (15%) and testing set (15%), which are used for training XGBoost model, selecting XGBoost structure, and testing the trained XGBoost model on new and unseen data points, respectively. The trained XGBoost models serve as the data-driven surrogate models for efficiency and ZVS. With the XGBoost models, the efficiency and ZVS performance under any unseen operating and modulation parameters can be evaluated.

### B. XGBoost Adopted in Stage I of the Proposed HEPS

In Stage I of the proposed HEPS modulation, XGBoost is adopted for building the data-driven models for ZVS and efficiency. XGBoost is a gradient-boosting-based ensemble learning approach, consisting of a bag of weak machine learning models such as decision trees [24].

The boosting training process of XGBoost models with $k$ decision trees is described with Fig. 9 for illustrative purposes. During the training of XGBoost, the $k$ decision trees are sequentially trained to fit the residual $o_i^*$, which is the difference between the objective value $o^*$ (such as $P_{loss}$, $n_{ZVS}$) and the cumulative sum of the outputs of all previous trees.

For instance, the prediction of Decision Tree 1 $y_1$ will try to follow $o_1^*$, and the residual $o_1^*-y_1$ will be the training objective $o_2^*$ for Decision Tree 2 to learn. Similarly, given $o_2^*$, Decision Tree 2 will output $y_2$ to get closer to $o_2^*$, and the corresponding residual $o_3^*$ (which is $o_2^*-y_2$) will be regarded as the training target for Decision Tree 3. This sequential training process of residuals is called bossting learning, and it is applied to all following trees. For Decision Tree $k$, its learnable parameters $\theta_k$ are adjusted to minimize the mean square error between its prediction $y_{k,\theta_k}$ and the $k^{th}$ residual $o_k^*$, as expressed in (4).

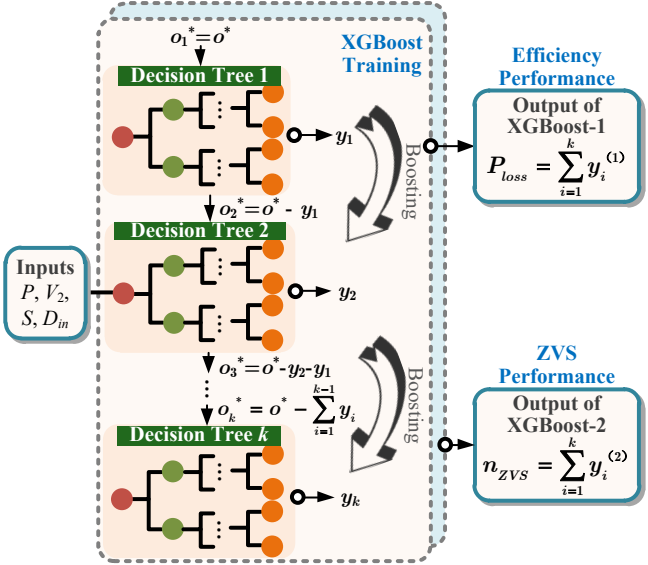

Fig. 9. Boosting training process of XGBoost.

$$\min_{\theta_k}(obj_k) = \min_{\theta_k}\left(o_k^* - y_{k,\theta_k}\right)^2 = \min_{\theta_k}\left(o^* - \sum_{i}^{k-1} y_i - y_{k,\theta_k}\right)^2 \quad (4)$$

$$P_{loss}(P,V_2,S,D_{in}) = \sum_{i=1}^{k} y_i^{(1)}(P,V_2,S,D_{in}) \quad (5)$$

$$n_{ZVS}(P,V_2,S,D_{in}) = \sum_{i=1}^{k} y_i^{(2)}(P,V_2,S,D_{in}) \quad (6)$$

The output of XGBoost models is obtained by the summation of the outputs of all decision trees. In this article, the output of model XGBoost-1 is the total power loss $P_{loss}$ as shown in (5), and XGBoost-2 predicts the number of switches that meet ZVS constraints $n_{ZVS}$ with (6). In (5) and (6), $y_i^{(1)}$ and $y_i^{(2)}$ are the outputs of the $i^{th}$ decision tree of XGBoost-1 model and XGBoost-2 model, respectively.

With the simulation data and the application of XGBoost in Stage I, data-driven models of ZVS and efficiency can be automatically built, and the challenges of time-consuming process and inaccuracy in conventional modeling approaches as discussed in Section III have been greatly conquered.

## V. STAGE II OF THE PROPOSED HEPS MODULATION: OPTIMIZE EFFICIENCY WITH FULL ZVS OPERATION

Stage II of the proposed AI-based HEPS modulation searches for the best EPS strategy and modulation parameters to realize optimal efficiency with full ZVS operation. Section V-A introduces the flowchart of Stage II, and Section V-B discusses the applied PSO-SAVL algorithm in Stage II.

### A. Stage II: Optimize Efficiency with Full ZVS Operation through PSO-SAVL Algorithm

This article aims to optimize efficiency while maintaining all-switch ZVS operation over entire voltage and power ranges, the mathematical form of which is expressed in (7) as the following:

**For the chosen operating parameters $S$, $P$ and $V_2$, the target is to minimize total power loss:**

$$P_{loss}^* = \min_{D_{in}}\left(P_{loss}(P,V_2,S,D_{in})\right), \quad (7a)$$

**Subject to:**

$$n_{ZVS}(P,V_2,S,D_{in}) = 8, \quad (7b)$$

$$0 \leq D_{in} \leq 1. \quad (7c)$$

In Stage II, the optimization problem in (7) is solved by the cutting-edge PSO-SAVL algorithm. Stage II is composed of three steps, the flowchart of which is given in Fig. 10. First, different values of $P$ and $V_2$ are chosen. Second, given the selected $P$ and $V_2$, a cutting-edge PSO variant, PSO-SAVL algorithm, is adopted to solve (7) for both $S$ = EPS1 and $S$ = EPS2. Afterwards, the optimal total power loss $P_{loss}^*$ of these two EPS strategies is compared, and the best EPS strategy under the specified $P$ and $V_2$ is selected.

Through the PSO-SAVL in Stage II, the best EPS strategy and the optimal $D_{in}$ can be obtained which can achieve the best efficiency and all-switch ZVS over the entire range of $P$ and $V_2$.

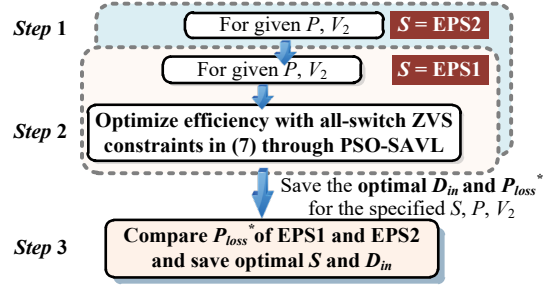

Fig. 10. Flowchart of Stage II of the proposed AI-based HEPS modulation.

### B. PSO-SAVL Adopted in Stage II of the Proposed HEPS

Particle swarm optimization (PSO) is a famous meta-heuristic optimization algorithm, which exhibits good performance in continuous problems [5]. PSO-SAVL, a state-of-the-art PSO variant, is chosen because of its robust global searching capability and fast convergence speed in low dimensional problems [27]. Different from many PSO variants, PSO-SAVL adaptively adjusts the velocity limit to match the evolutionary state of particles, the strategy of which boosts its holistic performance.

The flowchart of PSO-SAVL is given in Fig. 11 and is discussed as follows. Under the specified operating parameters of $S$, $P$ and $V_2$, PSO-SAVL are initialized first, such as velocity limit factors $vl_{min}$, $vl_{max}$, and learning factors $c_1$, $c_2$. Position $D_{in,j}$ and velocity $V_j$ of all particles are also initialized. Afterwards, based on the surrogate models of ZVS and efficiency from Stage I, the objective value $func_j$ is evaluated with (8), and personal best position $p_{best,j}$ and global best position $g_{best}$ are updated. In (8), $c_{ZVS}$ is the weight factor of ZVS constraint. Subsequently, evolutionary factor $f$ is evaluated with (9), where $d_{min}$ and $d_{max}$ are the minimum and maximum values of $d_j$, and $d_g$ is the $d_j$ of the globally best particle. Based on $f$, velocity limit $VL$ is updated with (10). Velocity $V_j$ of the $j^{th}$ particle is updated with (11), and the new $V_j$ is then constrained to the range [-$VL$, $VL$]. With the new and bounded $V_j$, position $D_{in,j}$ is updated and constrained within [0, 1]. This procedure repeats until the maximum epoch has been reached. If PSO-SAVL terminates, the optimal $D_{in}$ which has optimal efficiency and all-switch ZVS under the specified $S$, $P$ and $V_2$ can be found.

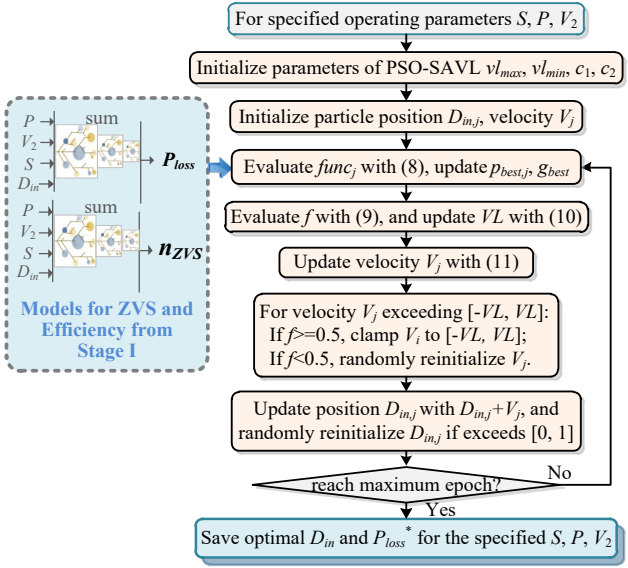

Fig. 11. Flowchart of PSO-SAVL applied in Stage II.

$$func_j = P_{loss} + \max(8 - n_{ZVS}, 0) \cdot c_{ZVS} \quad (8)$$

$$f = \frac{d_g - d_{min}}{d_{max} - d_{min}} \quad (9a)$$

$$d_j = \frac{1}{N-1} \sum_{k=1, k \neq j}^{N} |D_{in,j} - D_{in,k}| \quad (9b)$$

$$VL = \frac{1}{1 + \left(\frac{1}{vl_{min}} - 1\right) \exp\left(\ln\left(\left(\frac{1}{vl_{max}} - 1\right) / \left(\frac{1}{vl_{min}} - 1\right)\right) f\right)} \cdot D_{in,max} \quad (10)$$

$$V_j = \omega \cdot V_j + c_1 \cdot r_1 \cdot (pbest_j - D_{in,j}) + c_2 \cdot r_2 \cdot (gbest - D_{in,j}) \quad (11)$$

To sum up, under the selected $P$ and $V_2$, the best EPS strategy $S$ and the optimal inner phase shift $D_{in}$ can be searched by following the steps of Stage II.

## VI. DESIGN CASE WITH THE PROPOSED AI-BASED HEPS

In this section, by following the proposed AI-based HEPS approach, an efficiency-oriented hybrid EPS modulation with full ZVS range for DAB converters is designed.

### A. Stage I: Build Data-Driven Models of ZVS and Efficiency with XGBoost Algorithm

As a preparation for the proposed HEPS approach, the operating conditions of design case are specified in Table I, where the rated power $P_{rated}$ is 1000 W, and rated input and output voltages $V_{1,rated}$ and $V_{2,rated}$ are 200 V. $f_s$ is 20 $k$Hz and dead time is 400 $n$s. Leakage inductance $L_r$ is 167 $\mu$H as calculated by (3). In this design case, both buck and boost working conditions are considered: $V_2$ varies within [160 V, 240 V]. The adjustable range of $P$ is [100 W, 1000 W].

By following the flowchart of Stage I in Fig. 8, two equivalent data-driven models for ZVS and efficiency performance are built with XGBoost algorithm. In Step 1, 20×20×80=32,000 combinations of $V_2$, $P$ and $D_{in}$ are uniformly sampled for both EPS1 and EPS2. Since two EPS strategies are considered, the total number of simulations to run is 2×32,000=64,000. In Step 2, PLECS simulations with the chosen operating parameters are conducted to collect the total power loss $P_{loss}$ and the number of switches $n_{ZVS}$ that meet ZVS constraints. Finally, with the collected $P_{loss}$ and $n_{ZVS}$, two XGBoost models of ZVS and efficiency are trained, the configurations of which are shown in Table II.

TABLE I. SPECIFICATIONS OF DESIGN CASE

| Rated Conditions | | | |
|---|---|---|---|
| $P_{rated}$ | 1000 W | $V_{1,rated}$ | 200 V |
| $V_{2,rated}$ | 200 V | $f_s$ | 20 $k$Hz |
| Switching Device | | | |
| Device series | C2M0080120D | Dead time | 400 $n$s |
| Transformer | | | |
| Leakage inductor $L_r$ | | 167 $\mu$H | |
| Turn ratio $n$ | | 1 | |
| Ranges of Operating Parameters | | | |
| $P$ | [100 W, 1000 W] | $V_2$ | [160 V, 240 V] |

TABLE II. CONFIGURATIONS OF XGBOOST MODELS OF ZVS AND EFFICIENCY

| Inputs | $P$, $V_2$, $S$, $D_{in}$ |
|---|---|
| Learning rate | 0.08 |
| **Structure of XGBoost-1 for Efficiency Performance** | |
| Output | Total power loss $P_{loss}$ |
| Maximum tree depth | 9 |
| Regularization coefficient | 0.1 |
| Number of base tree models | 1930 |
| **Structure of XGBoost-2 for ZVS Performance** | |
| Output | Number of switches satisfying ZVS $n_{ZVS}$ |
| Maximum tree depth | 6 |
| Regularization coefficient | 1 |
| Number of base tree models | 189 |

### B. Stage II: Optimize Efficiency with Full ZVS Operation through PSO-SAVL Algorithm

With the flowchart of Stage II in Fig. 10, the best EPS strategy between EPS1 and EPS2 and the optimal $D_{in}$ are obtained for a broad range of $P$ and $V_2$. The parameters of the adopted PSO-SAVL in Stage II are given in Table III. The resulting optimal inner phase shift $D_{in}$ for strategies EPS1 and EPS2 are given in Fig. 12 and Fig. 13, respectively.

TABLE III. CONFIGURATIONS OF PSO-SAVL IN STAGE II

| Inputs | $P$, $V_2$, $S$, $D_{in}$, XGBoost-1 $P_{loss}(P, V_2, S, D_{in})$, XGBoost-2 $n_{ZVS}(P, V_2, S, D_{in})$ |
|---|---|
| Output | Optimal $D_{in}$ and optimal $P_{loss}^*$ for given $P$, $V_2$, $S$ |
| Number of particles | 5 |
| Maximum iterations | 50 |
| Weight inertia $\omega$ | Linearly decrease from 0.9 to 0.4 |
| ZVS weight factor | $c_{ZVS}$ = 100 |
| Learning factors | $c_1 = c_2 = 2.05$ |
| Velocity limit factors | $vl_{max}$ = 0.7; $vl_{min}$ = 0.4 |

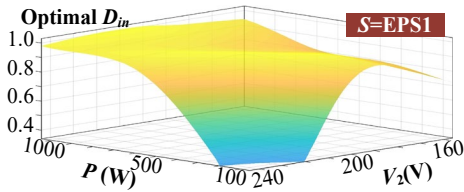

Fig. 12. Optimal $D_{in}$ for strategy EPS1 as obtained by Stage II.

With the optimal $D_{in}$, the ZVS ranges of EPS1 and EPS2 with respect to the considered ranges of $P$ and $V_2$ are shown in Fig. 14. For EPS1, it realizes all-switch ZVS operation in buck mode ($V_2 < 200$ V), but some switches lose their soft switching operation in boost mode ($V_2 > 200$ V) and light load situations. Conversely, EPS2 achieves ZVS for all 8 switches under boost situations, but it can only achieve partial ZVS in buck mode.

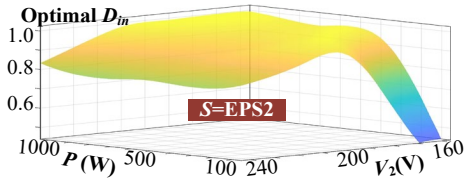

Fig. 13. Optimal $D_{in}$ for strategy EPS2 as obtained by Stage II.

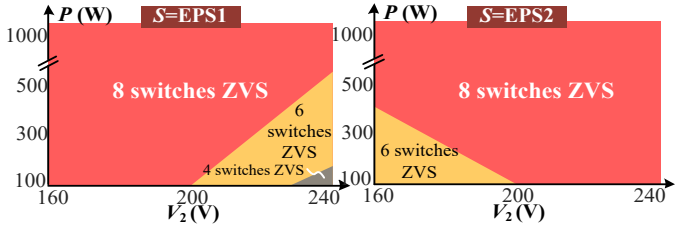

Fig. 14. ZVS ranges of EPS1 and EPS2 with the obtained optimal $D_{in}$.

The optimal efficiency performance for the two EPS strategies is shown in Fig. 15. 1000 W, 600 W and 200 W are taken as examples, and other power conditions share the same patterns as the following. As can be concluded from Fig. 15, in buck operating mode, EPS1 realizes higher efficiency than EPS2 for all three power levels, and EPS2 stands out in boost mode. Under the unit gain operating mode ($V_2 = 200$ V), from Figs. 12 and 13, $D_{in}$ of both EPS1 and EPS2 is optimized to 1, which is the same as SPS modulation.

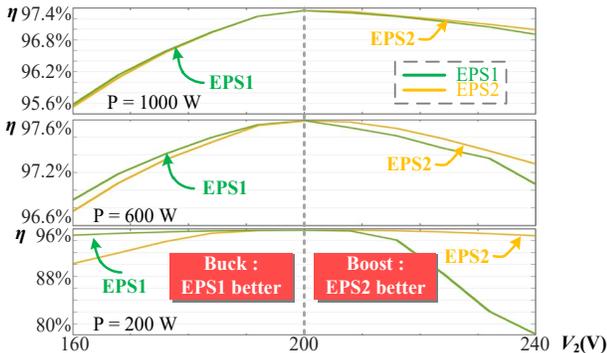

Fig. 15. Optimal efficiency of EPS1 and EPS2 with the obtained optimal $D_{in}$.

Consequently, in the proposed HEPS modulation, to realize optimal efficiency and all-switch ZVS operation over the entire operating ranges, EPS1 and EPS2 are applied in buck mode and boost mode, respectively. The optimal total power loss $P_{loss}^*$ of the proposed AI-based HEPS approach is given in Fig. 16. In HEPS modulation, the three-level voltage wave is applied to the high voltage side, reducing reactive power and circulating current, and thus contributing to high efficiency.

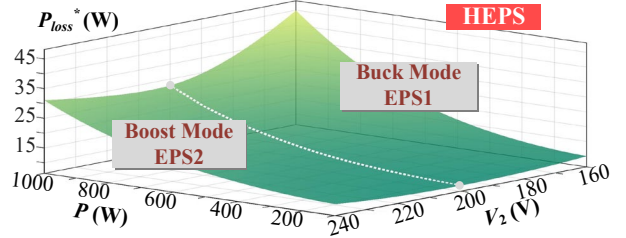

Fig. 16. Efficiency performance of the proposed AI-based HEPS modulation in the entire voltage and power ranges.

### C. Control Diagram of the Proposed HEPS Modulation

The control diagram of the proposed HEPS modulation is shown in Fig. 17. The output voltage $V_2$ is sampled to compare with the reference voltage value $V_{ref}$, and their difference $V_{ref} - V_2$ is fed into PI controller. To achieve closed-loop voltage and power regulation, $D_o$ is adjusted by PI controller. The hybrid EPS modulator is responsible for selecting the optimal EPS strategy based on the voltage conversion gain $M$ and tuning the inner phase shift given $V_{ref}$ and $P$. If it is the unit-gain mode, both $D_{in1}$ and $D_{in2}$ are 1, and the strategy simplifies to single phase shift modulation. In terms of the buck mode, optimal EPS1 strategy is adopted, and the optimal $D_{in1}$ is obtained from Fig. 12. If the situation is boost mode, optimal EPS2 strategy with $D_{in2}$ in Fig. 13 will be used.

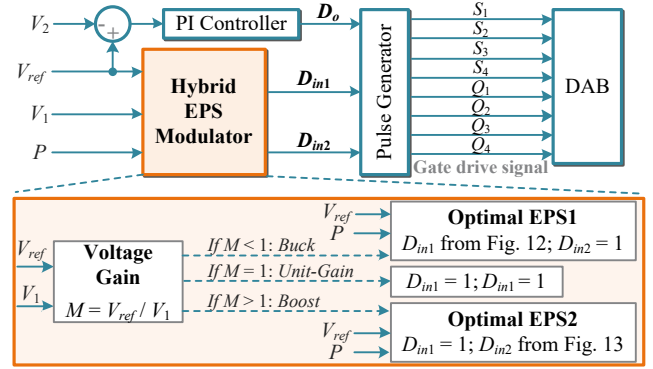

Fig. 17. Control diagram of the proposed AI-based HEPS modulation.

### D. Computational Time of the Proposed HEPS Modulation

To shed lights on the computational time required in the proposed AI-based HEPS modulation, under the computer platform with four-core Intel Xeon CPU E5-1630 @ 3.7 GHz, 16 GB RAM, and Windows 10 operating system, the average CPU time is given in Table IV.

Table IV. Computational Time of the Proposed HEPS Modulation

| Major Steps | Platform | Average CPU Time |
|---|---|---|
| Stage I: Run Simulation | Intel Xeon CPU E5-1630 @ 3.7 GHz, 16 GB RAM, 4 CPU cores, Windows 10 | 38 hours |
| Stage I: Build Data-Driven Models | | 4 minutes |
| Stage II: Optimization | | 56 minutes |

Based on Table IV, Stage I requires most of the computational time, where Step 2 takes 38 hours to run all 64000 times of simulation, and the training of XGBoost models in Step 3 only occupies few minutes. Compared to Stage I, the optimization process in Stage II requires much less time.

## VII. EXPERIMENTAL VALIDATION

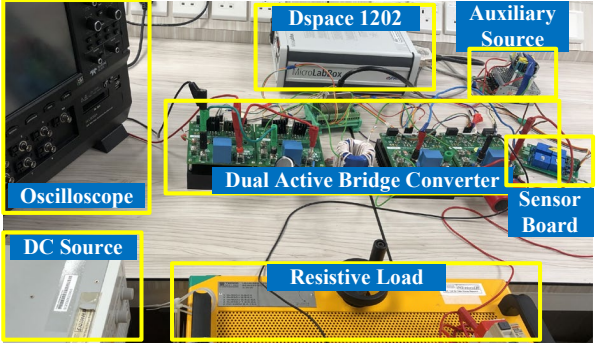

Fig. 18. Hardware platform for experimental validation.

To validate the design case in Section VI, hardware experiments have been carried out. Fig. 18 shows the designed hardware platform, which consists of an isolated DAB converter, a DC source, an auxiliary power supply, a variable resistive load, a DSPACE 1202 control platform, and a Lecroy oscilloscope. The specifications of the design case are listed in Table I.

### A. Experimental Waveforms

In this part, the steady-state experimental waveforms under different $P$ and $V_2$ are shown. The values of $P$ are 1000 W, 500 W and 100 W, which respectively represent high, medium and low power conditions. And $V_2$ can be 200 V, 160 V and 240 V, which respectively represent unit gain, buck and boost modes. The notations and directions of waveforms are given in Fig. 1.

Under the rated conditions when $P$ is 1000 W and $V_2$ is 200 V, the ac-side voltage waveforms $v_p$, $v_s$ are two-level waves as shown in Fig. 19, which is intrinsically the SPS modulation. The ZVS constraints of all 8 switches are satisfied, as the inductor current $i_L$ shown in Fig. 19 (b).

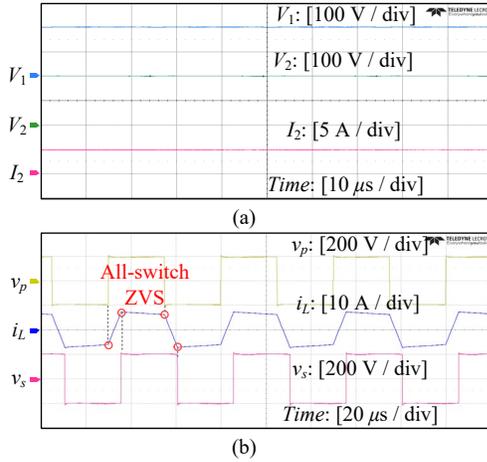

Fig. 19. Experimental waveforms under rated conditions when $P$ is 1000 W and $V_2$ is 200 V: (a) $V_1$, $V_2$ and $I_2$; (b) $v_p$, $i_L$ and $v_s$.

Under unit gain operation when $V_2$ is 200 V, the ac-side waveforms $v_p$, $i_L$ and $v_s$ of 500 W and 100 W are shown in Fig. 20. Figs. 19 and 20 together verify the all-switch ZVS operation under unit gain mode. Fig. 20 presents the waveforms of 1000 W, 500 W and 100 W under buck mode when $V_2$ = 160 V. When $V_2$ = 160 V, the optimal $D_{in}$ in Fig. 12 is applied to primary full bridge, which is EPS1 strategy, and all-switch ZVS requirements have been met. As shown in Fig. 22, for boost mode when $V_2$ = 240 V, EPS2 strategy with the optimal $D_{in}$ in Fig. 13 is selected, which achieves full ZVS operation over entire power range.

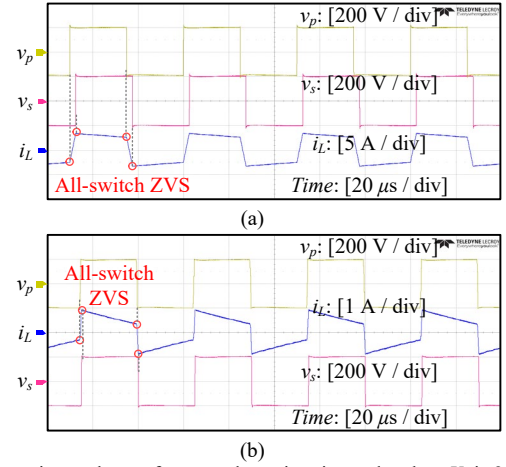

Fig. 20. Experimental waveforms under unit gain mode when $V_2$ is 200 V and: (a) $P$ = 500 W; (b) $P$ = 100 W.

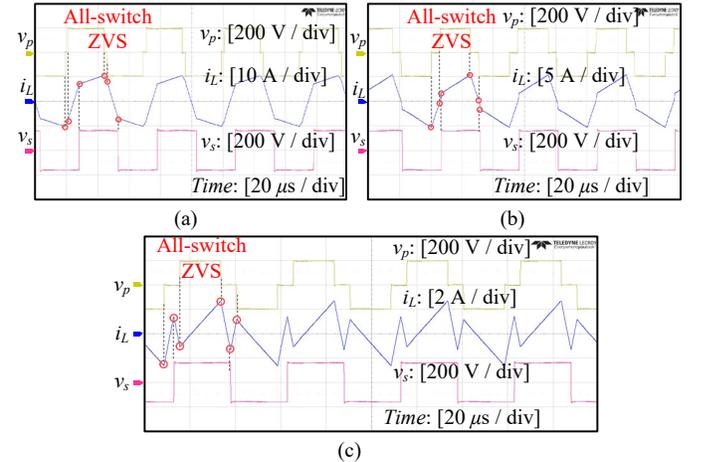

Fig. 21. Experimental waveforms under buck mode when $V_2$ is 160 V and: (a) $P$ = 1000 W; (b) $P$ = 500 W; (c) $P$ = 100 W.

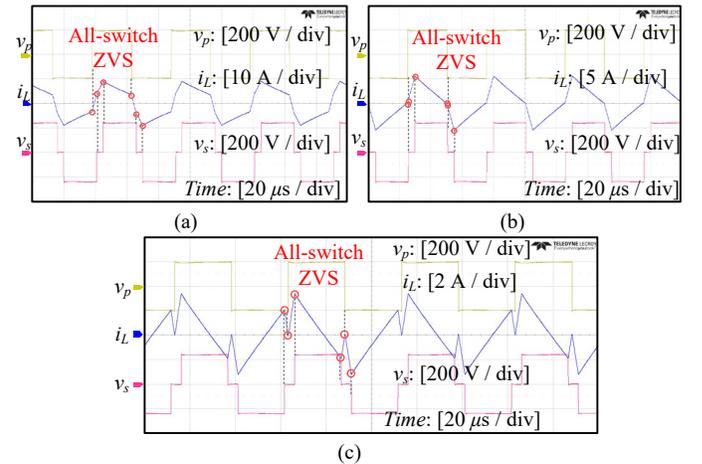

Fig. 22. Experimental waveforms under boost mode when $V_2$ is 240 V and: (a) $P$ = 1000 W; (b) $P$ = 500 W; (c) $P$ = 100 W.

### B. ZVS Analysis

To further validate the full ZVS operation of the proposed

HEPS modulation over the entire operating ranges, drain-source voltage $v_{ds}$ and gate-source voltage $v_{gs}$ of switches are given. Fig. 23 shows the ZVS waveforms when DAB converter operates under unit gain mode, where $v_{ds1}$ and $v_{gs1}$ are the $v_{ds}$, $v_{gs}$ waveforms of switch $S_1$, and $v_{ds5}$ and $v_{gs5}$ are the $v_{ds}$, $v_{gs}$ waveforms of switch $Q_1$. For load conditions of 1000 W, 500 W and 100 W, during the commutation process of all switches, $v_{ds}$ decreases to 0 V first, and then $v_{gs}$ rises to turn on the switches. Hence, under unit gain mode ($V_2$ = 200 V), all switches $S_1 \sim S_4$ and $Q_1 \sim Q_4$ have realized ZVS for the entire power range.

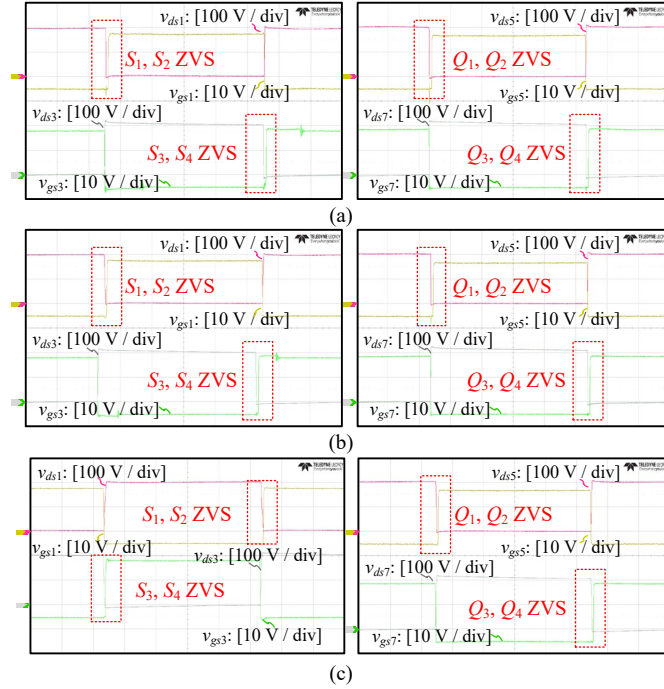

Fig. 23. ZVS waveforms $v_{ds1}$, $v_{gs1}$, $v_{ds3}$, $v_{gs3}$, $v_{ds5}$, $v_{gs5}$, $v_{ds7}$, $v_{gs7}$, when $V_2$ = 200 V and: (a) $P$ = 1000 W; (b) $P$ = 500 W; (c) $P$ = 100 W.

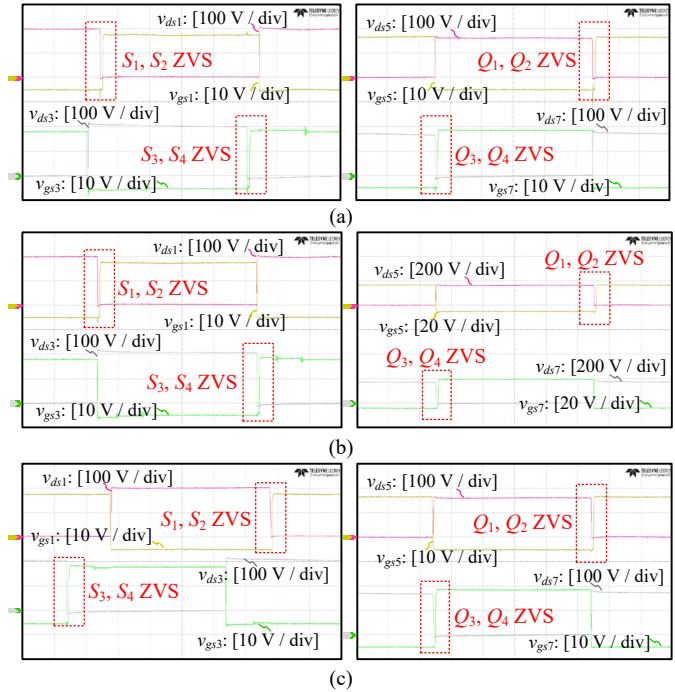

Fig. 24. ZVS waveforms $v_{ds1}$, $v_{gs1}$, $v_{ds3}$, $v_{gs3}$, $v_{ds5}$, $v_{gs5}$, $v_{ds7}$, $v_{gs7}$, when $V_2$ = 160 V and: (a) $P$ = 1000 W; (b) $P$ = 500 W; (c) $P$ = 100 W.

The ZVS waveforms $v_{ds1}$, $v_{gs1}$, $v_{ds3}$, $v_{gs3}$, $v_{ds5}$, $v_{gs5}$, $v_{ds7}$, $v_{gs7}$ of DAB converter under buck mode when $V_2$ is 160 V are shown in Fig. 24. There exists a gap between the falling of $v_{ds}$ and the rising of $v_{gs}$ for all switches, validating the full ZVS operation under buck mode. Moreover, Fig. 25 verifies the full ZVS operation under boost mode when $V_2$ is 240 V.

In summary, based on the steady-state waveforms in Figs. 19 to 22 and the ZVS waveforms in Figs. 23 to 25, for the entire power range and voltage range, the proposed HEPS modulation can realize full ZVS operation for all 8 switches.

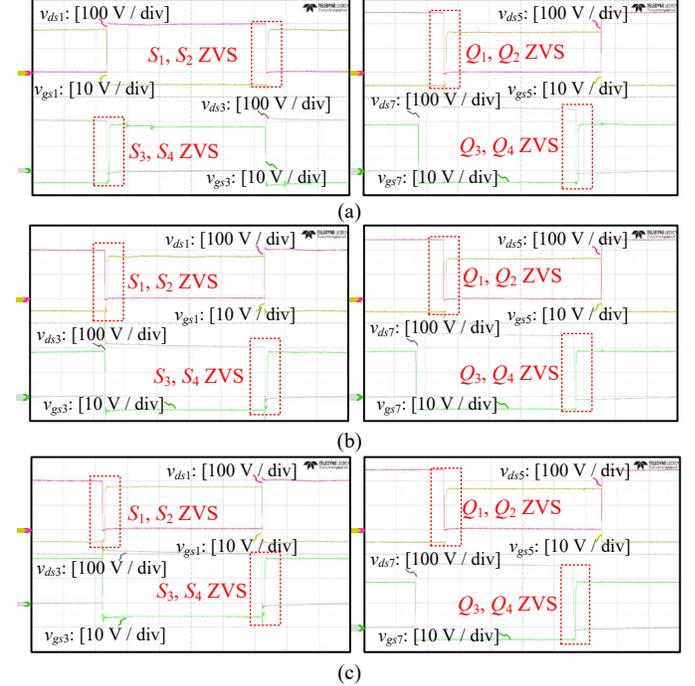

Fig. 25. ZVS waveforms $v_{ds1}$, $v_{gs1}$, $v_{ds3}$, $v_{gs3}$, $v_{ds5}$, $v_{gs5}$, $v_{ds7}$, $v_{gs7}$, when $V_2$ = 240 V and: (a) $P$ = 1000 W; (b) $P$ = 500 W; (c) $P$ = 100 W.

### C. Transient Response after Voltage and Load Step

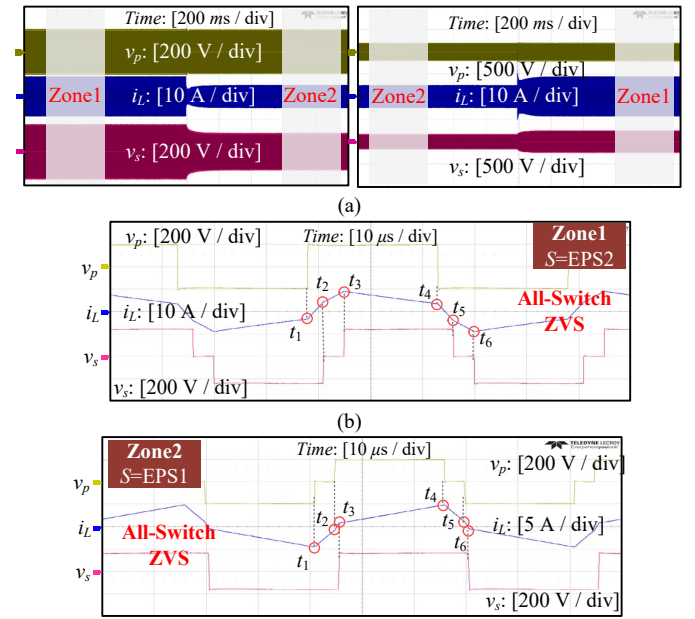

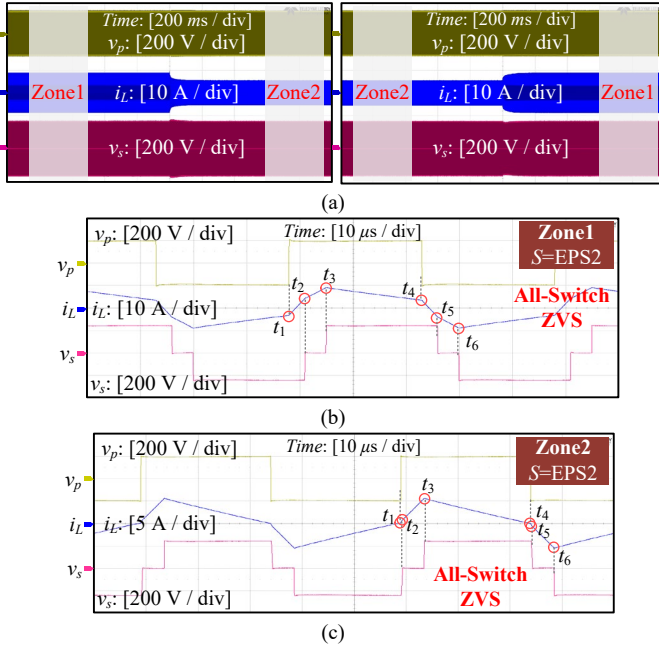

Fig. 26. Experimental waveforms during voltage steps when load resistance is 57.6 Ω: (a) $v_p$, $i_L$ and $v_s$ when $V_2$ steps from 240 V to 160 V and from 160 V to 240 V; (b) enlarged view of Zone1; (c) enlarged view of Zone2.

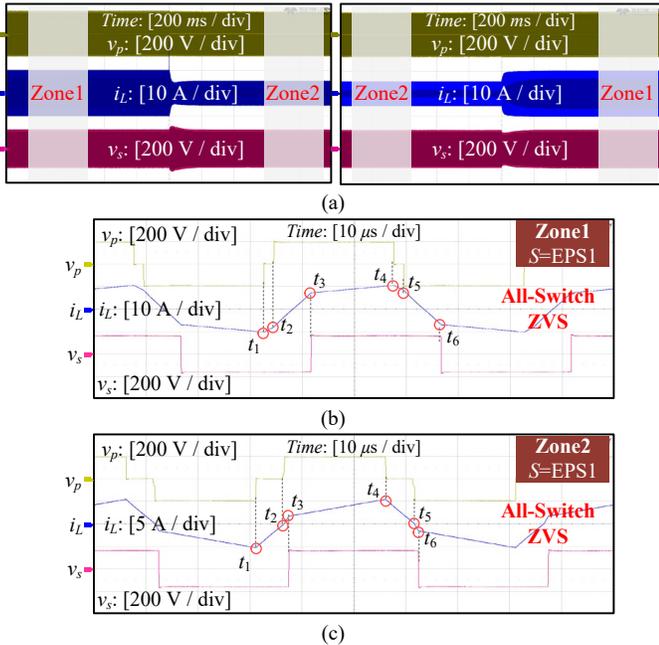

Fig. 27. Experimental waveforms during load steps when $V_2 = 240$ V: (a) $v_p$, $i_L$ and $v_s$ when $P$ steps from 1000 W to 500 W and from 500W to 1000 W; (b) enlarged view of Zone1; (c) enlarged view of Zone2.

Fig. 28. Experimental waveforms during load steps when $V_2 = 160$ V: (a) $v_p$, $i_L$ and $v_s$ when $P$ steps from 1000 W to 500 W and from 500W to 1000 W; (b) enlarged view of Zone1; (c) enlarged view of Zone2.

To verify that the proposed HEPS modulation can adaptively adjust the best EPS strategy and the optimal $D_{in}$ in varying operating conditions, the experiments of voltage and load step have been conducted in this part.

Under fixed load resistance of 57.6 Ω, the waveforms when $V_2$ steps between 240 V and 160 V are shown in Fig. 26. In Zone1 when $V_2$ is 240 V and $P$ is 1000 W, EPS2 strategy with the optimal $D_{in}$ is adopted, realizing all-switch ZVS as shown in Fig. 26 (b). As the zoom-in waveforms of Zone2 shown, EPS1 is utilized and all-switch ZVS operation is maintained. In the experiments of voltage step, the successful transition between two EPS modulations under different operating conditions validates the hybrid operation of the proposed HEPS modulation.

The experiments of load step under buck mode and boost mode are given in Fig. 27 and Fig. 28, respectively. When $V_2$ is 240 V and $P$ steps between 500 W and 1000 W, the best EPS modulation is EPS2 and the optimal $D_{in}$ is adjusted accordingly to achieve optimal efficiency and ZVS operation for all switches. As shown in Fig. 28, when $V_2$ is 160 V and load steps occur, EPS1 is the best strategy and $D_{in}$ has been tuned to its optimal.

In a nutshell, the experiments of voltage and load step comprehensively validate the real-time operation of the proposed AI-based HEPS modulation.

### D. Efficiency and ZVS Performance of the Proposed AI-based HEPS Modulation

The experimental performance of efficiency and ZVS of the proposed AI-based HEPS modulation is discussed as follows.

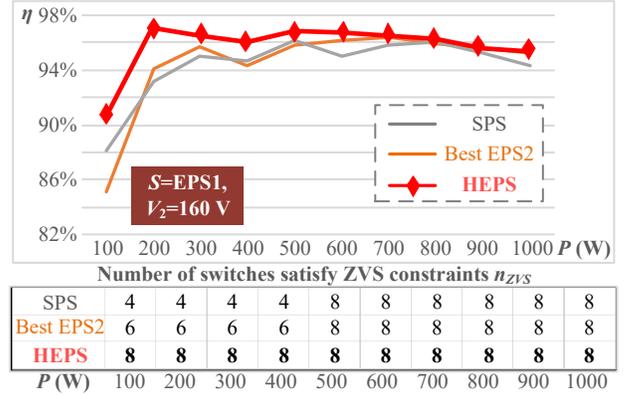

| SPS | 4 | 4 | 4 | 4 | 8 | 8 | 8 | 8 | 8 | 8 |
| Best EPS2 | 6 | 6 | 6 | 6 | 8 | 8 | 8 | 8 | 8 | 8 |
| HEPS | 8 | 8 | 8 | 8 | 8 | 8 | 8 | 8 | 8 | 8 |
| $P$ (W) | 100 | 200 | 300 | 400 | 500 | 600 | 700 | 800 | 900 | 1000 |

Fig. 29. Experimental efficiency ($\eta$) and ZVS performance ($n_{ZVS}$) of SPS, best EPS2, and the proposed HEPS approach in buck mode when $V_2 = 160$ V.

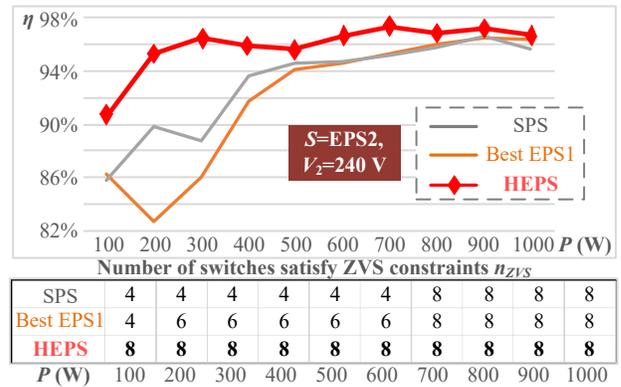

| SPS | 4 | 4 | 4 | 4 | 4 | 4 | 8 | 8 | 8 | 8 |
| Best EPS1 | 4 | 6 | 6 | 6 | 6 | 6 | 8 | 8 | 8 | 8 |
| HEPS | 8 | 8 | 8 | 8 | 8 | 8 | 8 | 8 | 8 | 8 |
| $P$ (W) | 100 | 200 | 300 | 400 | 500 | 600 | 700 | 800 | 900 | 1000 |

Fig. 30. Experimental efficiency ($\eta$) and ZVS performance ($n_{ZVS}$) of SPS, best EPS1, and the proposed HEPS approach in boost mode when $V_2 = 240$ V.

Under buck operating mode when $V_2$ is 160 V, efficiency $\eta$ and ZVS $n_{ZVS}$ of the proposed HEPS approach compared with other approaches are shown in Fig. 29, where the best EPS2 is the EPS2 strategy with the optimal $D_{in}$ in Fig. 13. The proposed HEPS strategy exhibits the best efficiency in the entire load range, and its peak efficiency is 97.1%. In terms of $n_{ZVS}$, the best

EPS2 strategy can improve the ZVS performance compared with SPS strategy, and our HEPS approach can realize all-switch ZVS over entire load range.

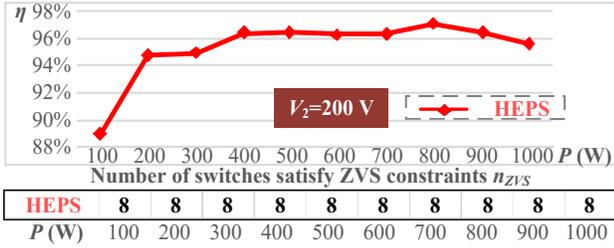

Fig. 31. Experimental efficiency ($\eta$) and ZVS performance ($n_{ZVS}$) of the proposed HEPS approach in unit gain mode when $V_2$ = 200 V.

Fig. 30 presents the performance in boost condition when $V_2$ is 240 V, where the best EPS1 is the EPS1 strategy with the optimal $D_{in}$ in Fig. 12. Compared with SPS and best EPS1 strategy, our HEPS strategy boosts efficiency by 4% when $P \leq$ 300 W, and it achieves the optimal efficiency in all load levels. From the perspective of ZVS performance, when $P$ is within [200 W, 600 W], the best EPS1 approach improves the 4-switch ZVS operation of SPS to 6-switch ZVS operation. The proposed HEPS approach can achieve 8-switch ZVS operation in all range.

The efficiency and ZVS performance of HEPS modulation in unit gain mode is given in Fig. 31. In unit gain mode, HEPS modulation is optimized to SPS strategy. Its peak efficiency reaches 97.08%, and all-switch ZVS operation is satisfied in the entire load range.

### E. Accuracy of the Theoretical Efficiency and ZVS Performance Evaluated by XGBoost Models

In this part, the theoretical efficiency and ZVS performance given by the trained XGBoost models is compared with experimental results, validating the high accuracy of the proposed AI-based data-driven approach. In terms of ZVS performance, Fig. 19 to Fig. 25 have validated the all-switch ZVS operation over the entire voltage and load ranges, which are aligned with the theoretical ZVS performance. From the perspective of efficiency, as shown in Fig. 32, the theoretical efficiency given by XGBoost-1 only has 0.67% deviations from experimental results in average. Based on the discussions above, it is reasonable to conclude that both XGBoost-1 and XGBoost-2 are aligned with the experimental results.

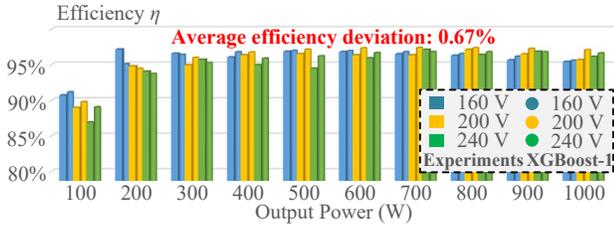

Fig. 32. Comparison between theoretical efficiency given by XGBoost-1 and experimental results.

In a word, the proposed AI-based HEPS modulation is experimentally validated to achieve the best efficiency while meeting all-switch ZVS operation over the full voltage and power ranges. Steady-state waveforms are presented in a comprehensive manner. Its full ZVS operation and its superior efficiency performance have been studied in detail. The experiments of voltage and load step verify the real-time operating capability of the HEPS approach. Besides, the comparison between the theoretical efficiency and ZVS performance and the experimental results validates the high accuracy of the data-driven XGBoost models.

### VIII. CONCLUSION

In this article, assisted by AI techniques, a hybrid extended phase shift (HEPS) modulation is proposed, which combines two EPS strategies. The proposed HEPS approach can optimize efficiency while maintaining all-switch ZVS operation over the full voltage and power ranges. With the integration of simulation software and XGBoost algorithm, the ZVS and efficiency models of the HEPS approach are developed in an automatic fashion, which alleviates the cumbersome and inaccurate model building process in conventional approaches. Generally, the proposed HEPS approach includes two stages. In Stage I, with the simulation-generated performance data, XGBoost algorithm is chosen to build data-driven models of ZVS and efficiency. Stage II utilizes the cutting-edge PSO-SAVL algorithm to optimize modulation strategy and parameter for the best efficiency and full ZVS operation. 1 kW hardware experiments on a DAB prototype validate the optimal efficiency, full ZVS range, real-time operating capability and thus the effectiveness of the proposed AI-based HEPS approach.

The proposed AI-based HEPS approach may suffer from mismatched data-driven models, leading to nontrivial deviations between theoretical performance and experimental results. Future research works can focus on solving the mismatched problem to ensure the accuracy of the trained data-driven models. A potential direction is to use simulation results completely for the training of models. Moreover, both simulation data and experimental results can be combined to train data-driven surrogate models. Besides, other advanced AI techniques such as few-shot learning and multi-task learning can be applied to ensure a good quality of data-driven models.

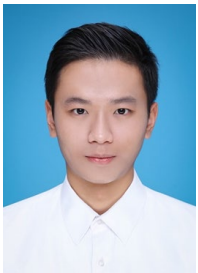

**Xinze Li** received his bachelor's degree in Electrical Engineering and its Automation from Shandong University, China, 2018. He has been awarded the Ph.D. degree in Electrical and Electronic Engineering from Nanyang Technological University, Singapore, 2023.

His research interests include dc-dc converters, modulation design, condition monitoring, digital twins for power electronics systems, design process automation, light and explainable AI for power electronics with physics-informed systems, application of AI in power electronics, and deep learning and machine learning algorithms.

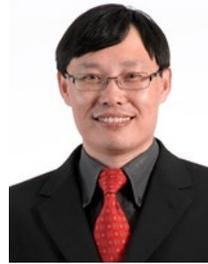

**Dr. Kezhi Mao** obtained his BEng, MEng and PhD from Jinan University, Northeastern University, and University of Sheffield in 1989, 1992 and 1998 respectively. He joined School of Electrical and Electronic Engineering, Nanyang Technological University, Singapore in 1998, where he is now a tenured Associate Professor.

Dr. Mao has over 20 years of research experience in artificial intelligence, machine learning, image processing, natural language processing, information fusion and cognitive science etc. He has published over 100 research papers in referred international journals and conferences. He has also edited 3 books published by Springer.

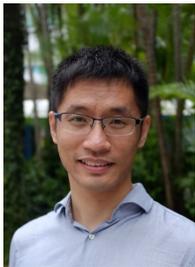

**Dr. Xin Zhang** received the Ph.D. degree in Automatic Control and Systems Engineering from the University of Sheffield, U.K., in 2016 and the Ph.D. degree in Electronic and Electrical Engineering from Nanjing University of Aeronautics & Astronautics, China, in 2014.

From February 2017 to December 2020, he was an Assistant Professor of power engineering with the School of Electrical and Electronic Engineering, Nanyang Technological University, Singapore. Currently, he is the professor at Zhejiang University. He is generally interested in power electronics, power systems, and advanced control theory, together with their applications in various sectors.

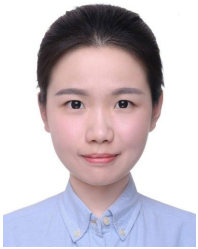

**Fanfan Lin** was born in Fujian, China in 1996. She received her bachelor degree in electrical engineering from Harbin Institute of Technology in China in 2018. From 2018, she has been awarded the Joint Ph. D. degree in Nanyang Technological University, Singapore and Technical University of Denmark, Denmark. Her research interest includes large language models for the design of large-scale power electronics systems, multi-modal AI for the maintenance of power converters, and the application of AI in power electronics.

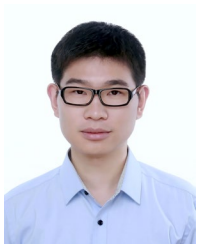

**Dr. Changjiang Sun** (S'13–M'19) received the Ph.D. degree from Shanghai Jiao Tong University, China, in 2019.

He is currently a Research Fellow with Nanyang Technological University, Singapore. His current research interests include topology and control of dc-dc converters for application in renewable energy collection and transmission grids.